\documentclass{eas}
\usepackage{graphicx}
\begin{document}

\title{Dark Matter Time Projection Chamber:  Recent R\&D Results} 
\author{J. B. R. Battat}\address{Physics Department, Bryn Mawr College; Bryn Mawr, PA 19010, USA}\email{jbattat@brynmawr.edu}
\author{S.~Ahlen}\address{Physics Department, Boston University; Boston, MA 02215, USA}
\author{M.~Chernicoff}\sameaddress{2}
\author{C.~Deaconu}\address{MIT Kavli Institute for Astrophysics and Space Research, Massachusetts Institute of Technology; Cambridge, MA 02139, USA}
\author{D.~Dujmic}\address{Laboratory for Nuclear Science, Massachusetts Institute of Technology; Cambridge, MA 02139, USA}
\secondaddress{Physics Department, Massachusetts Institute of Technology; Cambridge, MA 02139, USA}
\author{A.~Dushkin}\address{Physics Department, Brandeis University; Waltham, MA 02453, USA}
\author{P.~Fisher}\sameaddress{5} 
\author{S.~Henderson}\sameaddress{5}
\author{A.~Inglis}\sameaddress{2}
\author{A.~Kaboth}\sameaddress{5}
\author{L.~Kirsch}\sameaddress{6}
\author{J.P.~Lopez}\sameaddress{5}
\author{J.~Monroe}\address{Physics Department, Royal Holloway, University of London;  Egham, TW20 0EX, UK}
\author{H.~Ouyang}\sameaddress{6}
\author{G.~Sciolla}\sameaddress{6}
\author{H.~Tomita}\sameaddress{2}
\author{H.~Wellenstein}\sameaddress{6}
%
%
\runningtitle{J. B. R. Battat: DMTPC -- Recent R\&D Results}

\begin{abstract}
The Dark Matter Time Projection Chamber collaboration recently reported a dark matter limit obtained with a 10~liter time projection chamber filled with CF$_{4}$ gas.  The 10~liter detector was capable of 2D tracking (perpendicular to the drift direction) and 2D fiducialization, and only used information from two CCD cameras when identifying tracks and rejecting backgrounds.  Since that time, the collaboration has explored the potential benefits of photomultiplier tube and electronic charge readout to achieve 3D tracking, and particle identification for background rejection.  The latest results of this effort is described here.
\end{abstract}
\maketitle
\section{Introduction}
The Dark Matter Time Projection Chamber (DMTPC) collaboration recently reported their first limit on the spin-dependent interaction of Weakly Interacting Massive Particles (WIMPs) on protons using a 10~liter time projection chamber (TPC) filled with CF$_4$ (Ahlen \etal\ \cite{dmtpc}).  At that time, the DMTPC 10~liter detector was read out with CCD cameras only, and signal/background discrimination was based entirely on CCD image data.  The detector was capable of 2D tracking (no sensitivity parallel to the drift distance) and 2D detector fiducialization (in the same dimensions).  Motivated by the desire to improve particle identification, and to achieve full 3D tracking, the collaboration has explored the use of photomultipier tubes (PMTs) and electronic charge readout of various types.  
In particular, the status of track reconstruction and particle identification and vetoing is described.

\section{The DMTPC Detectors}
\label{desc}
Here we provide a brief description of the DMTPC technology to aid the following discussion.  Although different detectors were used for various aspects of the detector R\&D described here, many features are common across the detectors.  
In particular, the DMTPC technology makes use of a TPC filled with CF$_4$ gas.  Sometimes additional gases are used (e.g. $^4$He or $^3$He). 
The DMTPC technology is explained in detail elsewhere (see e.g. Ahlen \etal\ \cite{dmtpc} and Ahlen \etal\ \cite{whitepaper}).  
For the studies described here, typical drift distances of 20~cm were used and the amplification region diameter was $\sim$30~cm.  In all cases, the amplification region is imaged with at least one CCD (or EMCCD) camera to provide 2D particle tracking.  And, as described below, either charge readout electronics (charge-integrating and/or nanosecond rise-time amplifiers) or PMTs augmented the CCD readout.

We will refer to the $x$ and $y$ coordinates as the dimensions perpendicular to the drift direction, and we note that the CCD camera sensor exists in the $x$-$y$ plane.  We label the drift direction with the $z$-coordinate.  The drifting electrons arrive at an amplification region which consists of a stainless steel mesh separated from a copper anode by $\approx$0.5~mm.  The mesh is held near ground, and a large positive voltage (typically 600-700~V) is applied to the anode.  

\section{3D Track Reconstruction}
The CCD cameras image a 2D projection (in the $x$-$y$ plane) of a 3D track.  In principle, however, the extent of the track in the $z$-direction (referred to as $\Delta z$) is correlated with the temporal duration of an event as the ionized electrons arrive at the amplification region.  This idea has been pursued previously (see e.g. Fetal \etal\ \cite{fetal} for a demonstration of PMT-based 3D tracking of $\sim$5~MeV alpha particles from a $^{241}$Am source).  
Here, the $\Delta z$ reconstruction was performed in two ways:  with PMT readout, and through charge readout electronics.

\begin{center}
\begin{figure}
\includegraphics[height=0.32\textheight]{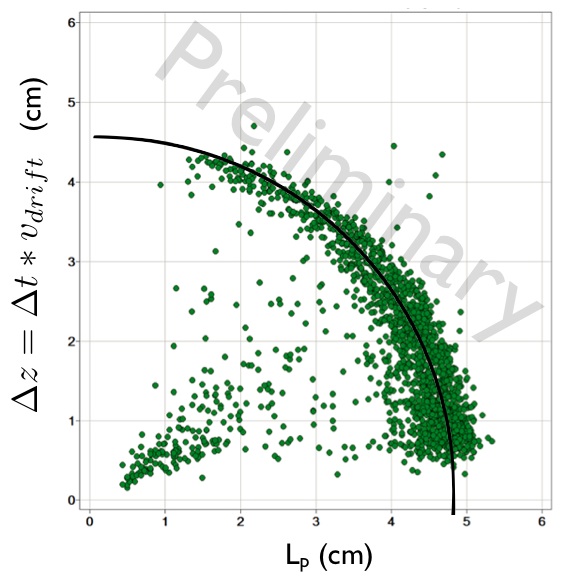}
\caption{Plot of the vertical length of a proton-triton track as inferred from the temporal width of the PMT signal, vs. the projected track length as measured with the EMCCD camera.  For the chosen drift velocity ($v_{drift}=0.87$~cm/s), the data lie along a quarter circle, as expected from Equation~\ref{l3d}.  Image taken from Tomita \cite{hideThesis}.}
\label{3dpmt}
\end{figure}
\end{center}

First we describe results using PMTs.  This work was carried out with the DMTPC Cylon detector, which is described in Chapter 8 of Tomita \cite{hideThesis}.  In addition to an EMCCD camera (Andor iXon +888), Cylon employs four PMTs (Photonis XP2017B) in four-fold coincidence.  The measured signal is the analog sum of the four PMTs.  Although the triggering efficiency as a function of recoil energy has not yet been fully characterized, we note that 20~keV events were seen.  To provide a sample of tracks of known length to use as a calibration sample for $\Delta z$ as reconstructed by the PMTs, Cylon was filled with the following gas mixture:  40~Torr CF$_4$ + 600~Torr $^4$He + 50~Torr $^3$He.  Helium-3 captures thermal neutrons which produces back-to-back emission of proton and triton particles:  $n^0+^3\mbox{He}\rightarrow P^+ + T^+ + 764\mbox{ keV}$.  Given the above gas mixture, the total track length of the pair is $\approx$5~cm.  A sample of proton-triton tracks were selected on energy, as measured by the EMCCD camera.  Then, the total track length, $L_{3D}$ was constructed from the projected length $L_{p}$ (measured with the EMCCD), and the temporal width $\Delta t$ of the PMT signal in the following way:
\begin{equation}
\label{l3d}
L_{3D} = \sqrt{L_p^2 + \left(\Delta t \times v_{drift}\right)^2}
\end{equation}
where $v_{drift}$ is the drift velocity of the electrons in the TPC (0.87 cm/$\mu$s for the gas mixture described above).  If the correct value of $v_{drift}$ is chosen, then a plot of $\Delta t \times v_{drift}$ vs. $L_p$ should produce the quadrant of a circle centered on the origin with radius $L_{3D}$ (in this case $\approx$5~cm).  In fact, the electron drift velocity can be measured in this way (by tuning $v_{drift}$ to minimize the scatter of the points).  Figure \ref{3dpmt} shows data from the Cylon detector to demonstrate that the temporal width of the PMT signal does correlate with $\Delta z$ and so the PMT data can be used to reconstruct the third dimension of particle tracks.

\begin{figure}
\includegraphics[height=0.3\textheight]{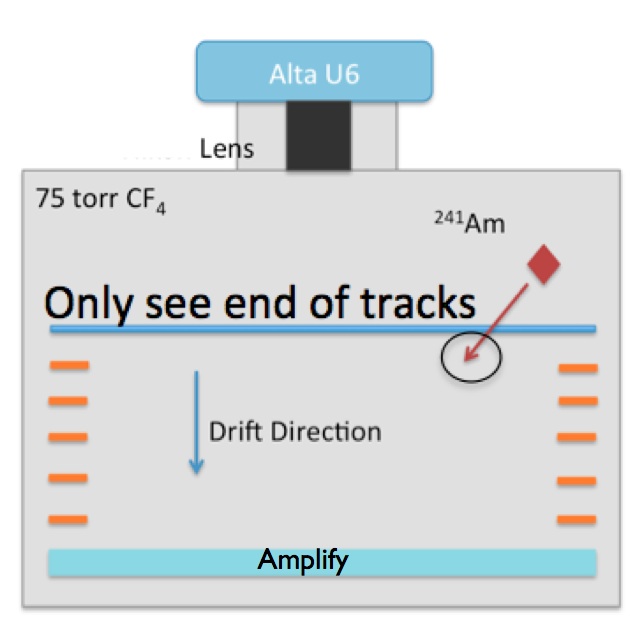}
\hfill
\includegraphics[height=0.3\textheight]{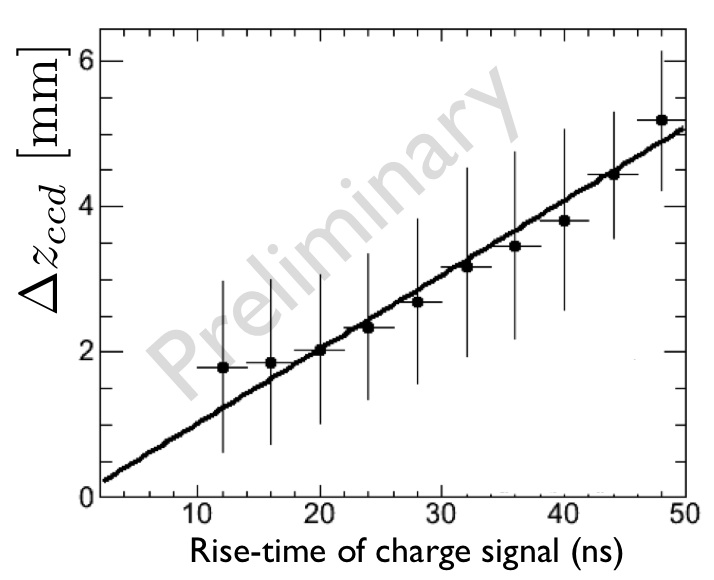}
\caption{Left:  Schematic showing the setup in which an $^{241}$Am is used to generate low energy particle tracks.  The source lies outside of the drift region and only the final $\approx$100~keV of the track is deposited in the detector's active region.  Right:  Correlation between $\Delta z_{CCD}$ and the rise-time of the charge signal.  See text for description of these quantities.}
\label{rndschematic}
\end{figure}

In addition to the PMT readout described above, the DMTPC collaboration has also explored the use of charge readout electronics to reconstruct $\Delta z$ for tracks.  By using a fast-timing amplifier (e.g. Ortec VT120 or Route2Electronics HS-AMP-CF) connected to the ground mesh of the amplification region (see Section \ref{desc}) it is possible to measure the rise-time of electronic pulses and correlate them with the $\Delta z$ of a track.  To produce low-energy tracks of known orientation, a $^{241}$Am alpha source is positioned in the vacuum chamber such that the majority of the energy loss occurs outside if the drift region (see Figure~\ref{rndschematic}).  Only the final $\approx$100~keV of the particle energy is deposited inside the drift region.  The angle of the source with respect to the $x$-$y$ plane is measured to be 27$^0$ below the horizontal.  The vertical extent of the track can therefore be calculated from the projected track length as imaged by the CCD camera in the following way
\begin{equation}
\Delta z_{CCD} = \tan(27^0)L_p\approx 0.5 L_p
\end{equation}
Figure~\ref{rndschematic} shows the observed correlation between the charge amplifier rise-time and $\Delta z_{CCD}$.  At the smallest values of $\Delta z_{CCD}$, the correlation weakens, largely due to difficulties in image-based track reconstruction at low energy.  Otherwise, the uncertainty in the reconstruction of $\Delta z$ through charge readout waveform timing is of order 1~mm.

\section{Background Rejection with Charge Readout}
There are several classes of CCD backgrounds that have no corresponding ionization event in the active region of the TPC.  For example, the interaction of a cosmic ray with the silicon of the CCD sensor.  The requirement that a coincident signal exist in a charge readout channel can significanly suppress these CCD backgrounds.  To demonstrate this, a DMTPC detector was operated without any radioactive sources present to study the number of background events that pass all CCD cuts vs. the number of events that pass CCD plus charge coincidence cuts.  For this experiment, the anode plane of the amplification region was segmented into a central circular region surrounded by an annular ``veto'' region.  Particles originating from the field cage or the walls of the vacuum chamber must deposit energy in this veto region and can subsequently be rejected.  The central anode was read out with a Cremat CR-113 charge integrating amplifier and the veto electrode was read out with a Cremat CR-112 amplifier.  Following a 14 live-hour exposure, 3399 events passed the CCD only cuts.  Once the charge-based coincidence cuts were applied (namely a coincident signal in the central anode and no signal in the veto electrode), only 12 events remained, corresponding to a CCD-background suppression of 99.6\%.  The range vs. energy of the remaining events was consistent with C or F nuclear recoils.

A related study was performed to measure the gamma-rejection rate achievable with charge readout electronics.  
Electronic recoils have a different time profile in charge than do nuclear recoils.  The latter are generally compact and therefore have a smaller $\Delta z$ than electronic recoils.  During amplification, the majority of the fast-moving electrons are created near the anode and form a peak that decays on the timescale of 1~ns.  A subsequent ion peak, caused by the ions that drift from the anode to the ground mesh, create a second, broader peak.  Unlike nuclear recoils, electronic recoils produce primary electron clouds over a larger range of $z$ which results in a single, broad amplification peak in the charge signal.  

To evaluate the gamma and electron rejection capability, a 5~$\mu$Ci $^{137}$Cs source was placed above the cathode mesh, pointed along the $z$-axis toward the amplification region.  During a 14.2 hour exposure, an estimated 409800 electronic recoils with 40 keV$_\alpha < E < 250$~keV$_\alpha$ occurred in the fiducial region defined by the field of view of the CCD.  Of those, five events pass the combined charge and CCD cuts, and are consistent with being nuclear recoils in terms of both range vs. energy and maximum track CCD pixel intensity.  Therefore, the number of electronic recoils seen by the CCD is zero, when 409800 were present.  This corresponds to a 90\% C.L. upper limit on the electronic rejection factor of 5.6$\times$10$^{-6}$, and is statistics limited.  For a detailed discussion of this work, see Lopez \etal\ \cite{lopez}.  

\begin{figure}
\includegraphics[height=0.3\textheight]{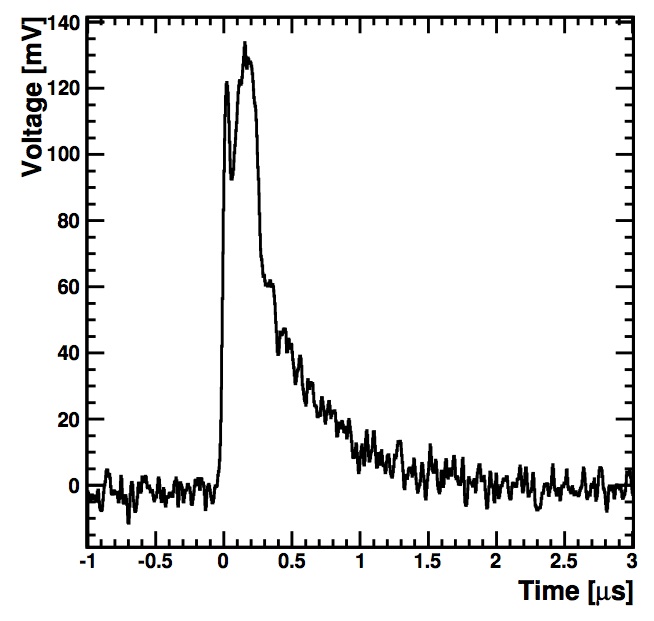}
\hfill
\includegraphics[height=0.3\textheight]{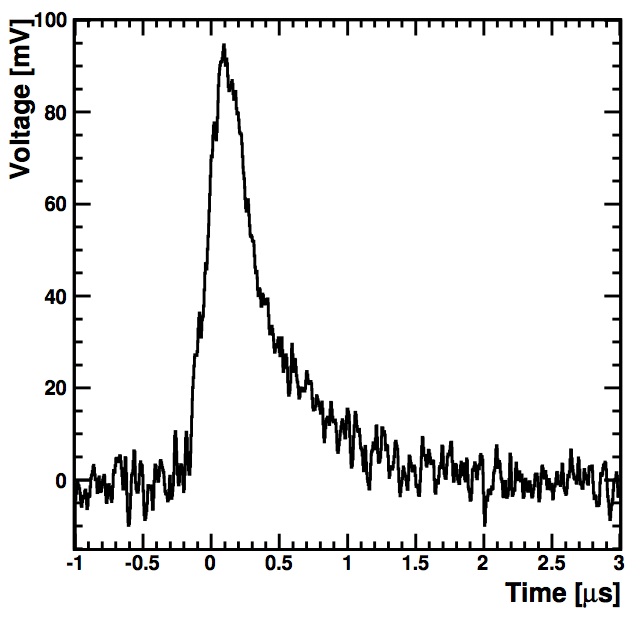}
\caption{Waveforms from the Route2Electronics HS-AMP-CF amplifier connected to the ground mesh of the amplification region.  Left:  A nuclear recoil produces two distinct peaks:  the narrow electron peak and the broader ion peak.  Right:  An electronic recoil produces a single, broad peak because the primary electrons were created over a large range of $z$.}
\label{meshamp}
\end{figure}

\section{Conclusion}
Following the recent dark matter surface run (Ahlen \etal\ \cite{dmtpc}), the DMTPC collaboration has explored techniques to reconstruct the third track dimension and to enhance the background rejection capability.  Through the use of both PMTs and fast charge amplifiers, we have demonstrated sensitivity to the $\Delta z$ of a track.  In addition, we demonstrated enhanced background rejection capability using charge readout in coincidence with CCD imaging.




\end{document}